\documentclass{article}
\usepackage{spconf,amsmath,graphicx,hyperref}
\usepackage{cite}
\usepackage{amssymb,amsfonts}
\usepackage{algorithm}
\usepackage{algorithmic}
\usepackage{textcomp}
\usepackage{pgfplots}
\usepackage{xcolor}
\usepackage{arydshln}
\usepackage{tikz}
\usepackage{pgfplots}
\usepackage[caption=false,font=footnotesize]{subfig}
\pgfplotsset{compat=1.18}

\definecolor{OIOrange}{HTML}{E69F00} % *AudioSeal
\definecolor{OIBlue}  {HTML}{0072B2} % BloodRoot
\definecolor{OIGreen} {HTML}{009E73} % JingleBack
\definecolor{OIPink}  {HTML}{CC79A7} % Ultrasonic
\definecolor{OIGray}  {gray}{0.15}   % PBSM (dark gray)

\title{Bloodroot: When Watermarking Turns Poisonous for Stealthy Backdoor}
\name{Kuan-Yu Chen$^{1,2}$, Yi-Cheng Lin$^{1}$, Jeng-Lin Li$^{2,\star}$, Jian-Jiun Ding$^{1,\star}$}

\address{
$^{1}$Graduate Institute of Communication Engineering, National Taiwan University\\
$^{2}$AI Research Center, Inventec Corporation\\  
}

\begin{document}
\ninept
\maketitle

\begin{abstract}
Backdoor data poisoning is a crucial technique for ownership protection and defending against malicious attacks. Embedding hidden triggers in training data can manipulate model outputs, enabling provenance verification, and deterring unauthorized use. However, current audio backdoor methods are suboptimal, as poisoned audio often exhibits degraded perceptual quality, which is noticeable to human listeners. This work explores the intrinsic stealthiness and effectiveness of audio watermarking in achieving successful poisoning. 
%We propose Bloodroot backdoor framework to improve perceptual quality through adversarial LoRA finetuning, while maintaining a high trigger success rate and strong clean-sample accuracy.
We propose a novel Watermark-as-Trigger concept, integrated into the Bloodroot backdoor framework via adversarial LoRA fine-tuning, which enhances perceptual quality while achieving a much higher trigger success rate and clean-sample accuracy. Experiments on speech recognition (SR) and speaker identification (SID) datasets show that watermark-based poisoning remains effective under acoustic filtering and model pruning. The proposed Bloodroot backdoor framework not only secures data-to-model ownership, but also well reveals the risk of adversarial misuse.

%Audio triggers play a critical role in the privacy of speech signals. Applying hidden triggers to the training data can cause speech signals to be unable to be identified by a recognition model. It can ensure rightful ownership, verify provenance, and prevent unauthorized use. However, many existing triggers are easy to notice. Moreover, they may lower the audio quality, and user information is hard to retrieve, which limits their use in real-world ownership verification. In this work, we use audio watermarking as a stealthy and owner-specific trigger. By taking advantage of watermarking’s ability to stay hidden and keep audio clear, the proposed method creates triggers that are hard to detect, unique to each owner, and effective at activating backdoors. Experiments on benchmark datasets show high attack success rates while maintaining audio quality and enabling ownership tracing.
\end{abstract}

\vspace{-1em}
\section{Introduction}
\label{sec:intro}
\vspace{-4pt}

Deep neural networks (DNNs) have achieved widespread success in various speech applications, including speech recognition (SR) and speaker identification (SID)~\cite{dynamic-superb}. 
The non-transparent usage of training data for the speech models raises significant ownership concerns~\cite{Dataownership}. Backdoor data poisoning serves as a viable approach to enable detectable evidence for ownership protection. The data owner could inject a specific \emph{poison trigger} into a part of the training samples. The trained model consequently changes its prediction behaviors while inferring on the samples with the trigger~\cite{backdoor-survey}. 
For example, a speaker is incorrectly identified as another speaker once the backdoor is triggered~\cite{poison-sr,poison-speakerid}. Similarly, the SR victim model is triggered to output only a specific decoding token while maintaining decoded results in clean input cases~\cite{Backdoor-stealthy}.
Backdoor poisoning can also be exploited as a malicious attack during the training stage. %In our setting, the adversary has no access to the victim model or its training process and can only compromise the training data itself.

Previous works have made rapid progress in the dedicated crafting of audio backdoor triggers, such as timbre or pitch modification~\cite{JingleBack,Backdoor-stealthy}, ultrasonic pulse insertion~\cite{Ultrasonic}, and ambient sound mixing~\cite{Backdoor-ACM}. More recently,  some training strategies have integrated audio compression, taking advantage of its natural alignment while preserving imperceptible differences in audio quality~\cite{CBA}.
Although several methods claim a promising attack success rate, added poisons still lack sufficient usability due to two challenges:
(i) \emph{Perceptual quality}: Most poisons introduce acoustic artifacts, which are noticeable to humans.
(ii) \emph{Poison robustness}: Poisoned audio samples may be intrinsically exposed to various pre-processing and post-training procedures, accidentally eliminating the poison pattern and causing trigger failure.
Most poisoning studies focus primarily on malicious objectives, yet leave room for further improvements in perceptual quality and robustness against common defenses.

%(iii) \emph{Poisoned data attribution}: poisons are implemented without data attribution, restricting the controllability for ownership protection purposes.
%Most poisoning studies focus primarily on malicious objectives, yet leave room for further improvements in perceptual quality and poison robustness. Moreover, they overlook the need for data attribution in ownership protection, which stands in contrast to the concealment goals of malicious attackers.

%\medskip
%\noindent\textbf{Limits of existing audio triggers.}
%Previous work has explored several trigger techniques in audio: timbre modification (e.g. effect chains)\cite{JingleBack}, ultrasonic cues\cite{Ultrasonic}, dual adaptive augmentation~\cite{Backdoor-ACM}, pitch/voiceprint manipulation~\cite{Backdoor-stealthy}, and compression-based design~\cite{CBA}. 
% Although many methods achieve promising attack success, three issues remain.
% (i) \emph{Perceptual quality}: Even subtle perturbations can reduce intelligibility and lower PESQ, harming user experience.
% (ii) \emph{Detectability}: Many triggers leave spectral or temporal artifacts that listeners or automated defenses expose (e.g., filtering, pruning, activation clustering)~\cite{Backdoor-stealthy}.
% (iii) \emph{Real-world robustness and attribution}: Robustness against post-processing and channel effects (compression, filtering, resampling, replay) is limited, and most designs are not owner-specific, complicating accountability.

Due to the need to maintain audio quality while embedding poisoning patterns deeply and robustly, we advocate the use of audio watermarking techniques to encode imperceptible and enduring patterns inherently designed in watermark network pretraining~\cite{Watermark-poison,audioseal}. 
This repurposes audio watermarking as a backdoor trigger and raises an underexplored question: Can the imperceptible and robust properties of watermarking be sustained after victim models are trained?
In this work, we adopt a \emph{poison-only and trigger-based} setting: A small fraction of poisoned samples is used in model training \emph{without} modifying the training code or the system architecture. 
During inference, the trigger can activate the victim model to manipulate its prediction. 
Our setting is highly concerned with realistic implementations, which is similar to the setting of clean-label attacks~\cite{clean-label}.

We propose \textbf{Bloodroot}, a framework that repurposes audio watermarking as backdoor triggers for speech systems. 
Rather than a single model, Bloodroot provides a systematic way to embed imperceptible and robust watermark patterns into training data, enabling stealthy and effective triggers under realistic poisoning conditions. 
Within this framework, \textbf{Bloodroot} refers to AudioSeal without fine-tuning, while \textbf{Bloodroot-FT} applies LoRA finetuning, providing stronger imperceptibility and a higher attack success rate.

The SR and SID experiments demonstrate 32.5\% and 18.5\% relative perceptual evaluation of speech quality (PESQ) improvements while maintaining over 95\% success rate. Further analyses illuminate the poison robustness across network structures, signal filtering, and model pruning.  
In a nutshell, our contributions are as follows. 
\vspace{-2pt}
\begin{itemize}
\setlength{\itemsep}{-2pt}
    \item \textbf{Watermark-as-trigger framework}: We present the first approach that systematically uses audio watermarking as backdoor triggers. With the advanced audio watermarking model (e.g., LoRA-finetuned AudioSeal), a robust trigger with high imperceptibility can be designed. 
    
    \item \textbf{Backbone enhancement}: In addition, durable poisoning patterns are adopted to achieve high perceptual quality and minimal impact on benign accuracy. 

    \item \textbf{Extensive validation}: We conducted experiments on SR and SID tasks across diverse datasets and further assessed resilience against common defenses such as signal filtering and model pruning, where the proposed watermark-based triggers remain effective while conventional triggers fail.
   
    %\item \textbf{Comprehensive evaluation}: We evaluate the framework on SR and SID tasks across multiple datasets and model architectures, showing its wide applicability. 
    
    %\item \textbf{Robustness validation}: We test robustness against common defenses such as signal filtering and model pruning, where watermark-based triggers maintain effectiveness while conventional triggers collapse.
\end{itemize}
\vspace{-2pt}
While watermarking techniques were originally designed for ownership verification, our results show that their imperceptible and robust nature can be exploited as powerful poisoning triggers. 
This dual-use property underscores the importance of studying watermark-as-trigger attacks, including their risks and the way to design stronger defenses in this adversarial paradigm. Code is available at \href{https://github.com/DanielChen1128/Bloodroot-Audio-Backdoor}{GitHub}.
%The enhanced stealthiness and attribution of audio watermarks also enable end-to-end data-to-model ownership protection in a unified framework. 
%Notably, malicious use of watermarked models from another entity can enable powerful poisoning attacks and mislead source attribution, highlighting the need to advance defenses in this new adversarial paradigm.

%\noindent Together, these results bridge backdoor stealthiness and ownership verification. The proposed watermark-based triggers offer a practical path to stealthy backdoors with attribution, informing both attack understanding and the design of future defenses for speech systems.

\vspace{-1em}

\vspace{-1em}
\begin{figure}[t]
  \centering
  \includegraphics[width=1\columnwidth]{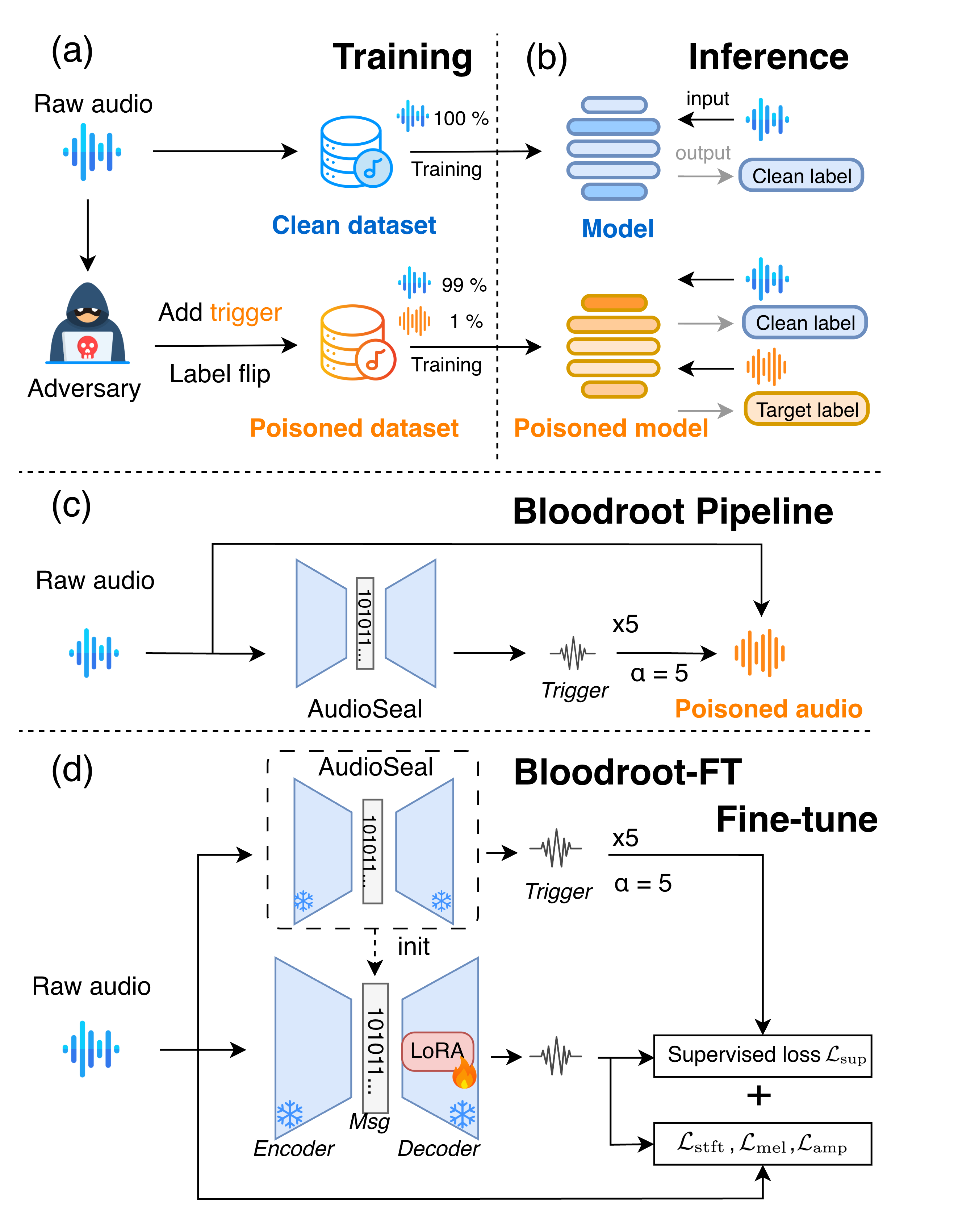}
  \vspace{-3.5em} % 建議稍微調整，避免過度擠壓
  \caption{
    Overview of the backdoor attack and Bloodroot framework.
    \textbf{(a) Training:} A victim model is trained on a dataset containing a small fraction of poisoned samples. 
    \textbf{(b) Inference:} Triggered inputs activate the backdoor (targeted misclassification), while clean inputs are processed normally. 
    \textbf{(c) Bloodroot:} The base attack uses a pre-trained AudioSeal generator; \textbf{``x5''} denotes a poison level of $\alpha=5$ to scale the trigger perturbation. 
    \textbf{(d) Bloodroot-FT:} LoRA fine-tuning refines the generator to optimize the trade-off between robustness and imperceptibility.
  }
  
  %Shield mark optional defenses (filtering and pruning).}
  \label{fig:pipeline}
\end{figure}

\vspace{-3em}

\section{Method}
\label{sec:method}

\vspace{-1.5pt}
\subsection{Poisoning Setup and Pipeline}
\label{sec:method:attack}
The overall poisoning pipeline is shown in Figure~\ref{fig:pipeline}, comprising (a) the training stage, (b) the inference stage, and (c,d) the proposed poison generation stages.
We consider a practical threat model where the adversary lacks access to the victim's training pipeline and internal parameters. To bypass detection, only a small portion (e.g., 1\%) of the training data is tampered with, reflecting real-world risks in large-scale dataset collection. Instead of designing complex noise patterns, we propose \textbf{Bloodroot}, a framework that repurposes existing audio watermarking models as effective poisoning triggers. The adversary embeds imperceptible watermark perturbations into a subset of training samples while assigning them to a target label. This ensures that poisoned samples remain indistinguishable from clean data. Once the victim trains on this compromised dataset, the resulting model behaves normally on clean inputs but yields targeted incorrect predictions whenever the watermark trigger is present. To further enhance the attack, we introduce \textbf{Bloodroot-FT}, where the adversary fine-tunes the watermark generator prior to poisoning to improve the trigger's robustness and imperceptibility.

\subsection{Audio Watermarking and Poisoning Properties}
\label{sec:method:wm}
% \vspace{-1em}
Learning-based audio watermarking methods have been developed in recent years~\cite{liu2025xattnmark, wavmark, audioseal, timbreMark, TraceableSpeech}. Typically, watermarks are generated with pretrained neural codecs~\cite{Encodec}. They focus on two key properties: \textbf{imperceptibility}, which is to be inaudible to human listeners, and \textbf{robustness}, which is to remain verifiable after resampling, compression, noise, or equalization. We exploit these pretrained features to strengthen the effectiveness of the backdoor. 

\textit{Backdoor attacks} are typically implemented using fine-grained audio patterns embedded in spectrograms. Their design objectives include effectiveness, stealthiness, and persistence~\cite{Backdoor-stealthy}. 
Although prior methods achieve high attack success rates, they often degrade audio quality. Therefore, amplifying poisoning effects while suppressing extraneous artifacts remains a challenging task. We hypothesize that the concealability of audio watermarks can help to hide backdoor patterns. Moreover, audio watermarking naturally enables attribution in the backdoor context. However, its latent capacity was overlooked in previous studies on audio poisoning.

\vspace{-1em}
\subsection{Watermark-to-Trigger Poison Generation}
\label{sec:method:adapt}

We propose \textbf{Bloodroot}, a backdoor framework that leverages audio watermarking as poisoning triggers, based on the observation that watermarks can induce poisoning effects when properly embedded. We use the AudioSeal model~\cite{audioseal} as a trigger generator to insert watermark patterns into targeted audio samples and later activate victim models with the same signals. Empirically, AudioSeal pretrained in VoxPopuli~\cite{voxpopuli} produces poisoned samples with high perceptual quality (PESQ), indicating strong stealth potential. However, because watermarking was not designed for backdoors, its performance in poisoning is suboptimal. To address this, we explore lightweight finetuning of the AudioSeal-based generator to improve attack effectiveness while retaining imperceptibility.

\begin{algorithm}[t]
\caption{Generalized Poisoning with Watermark-as-Trigger}
\label{alg:poison}
\begin{algorithmic}[1]
\REQUIRE Training set $\mathcal{D}=\{(x_i,y_i)\}_{i=1}^{N}$; watermark generator $G_{\alpha}(\cdot)$ ($\alpha$ can control the poison level; target label $y_t$
\ENSURE Poisoned dataset $\mathcal{D}'$ with $|\mathcal{D}'|=|\mathcal{D}|$
\STATE $\mathcal{I}_{\mathrm{non}} \gets \{\, i \mid y_i \neq y_t \,\}$ \COMMENT{indices of non-target samples}
%\STATE $K \gets \lfloor \rho \cdot N \rfloor$
\STATE Select $\mathcal{P} \subset \mathcal{I}_{\mathrm{non}}$ with $|\mathcal{P}|=\rho N$ uniformly at random (w/o replacement and $\rho$ is the poisoning rate)
\STATE Initialize $\mathcal{D}' \gets \emptyset$
\FOR{$i=1$ \TO $N$}
  \IF{$i \in \mathcal{P}$}
    \STATE $w_i \gets G_{\alpha}(x_i)$ \COMMENT{generating watermark}
    \STATE $\tilde{x}_i \gets x_i + w_i$ \COMMENT{embed trigger; no extra scaling here}
    \STATE $y_t = \tilde{y}_i \gets y_t$ \COMMENT{label-flip to target; using $y_i$ if clean-label}
    \STATE Add $(\tilde{x}_i, \tilde{y}_i)$ to $\mathcal{D}'$ \COMMENT{replacing original $(x_i,y_i)$}
  \ELSE
    \STATE Add $(x_i, y_i)$ to $\mathcal{D}'$ \COMMENT{keeping clean sample}
  \ENDIF
\ENDFOR
\RETURN $\mathcal{D}'$
\end{algorithmic}
\end{algorithm}

\begin{table*}[!htbp]
\centering
\caption{\textbf{Performance of the proposed Bloodroot and baseline backdoor attacks on keyword spotting tasks (SC-10 and SC-30) at the 1\% poison rate}. 
BA: benign accuracy (\%); ASR: attack success rate (\%). Higher PESQ and STOI indicate better perceptual quality and intelligibility. The
\textbf{Bold} style indicates the best performance, and the \underline{underlined} style indicates the second best.}
\label{tab:sr-results}
\renewcommand{\arraystretch}{1.1}
\setlength{\tabcolsep}{6.5pt}
\begin{tabular}{l|cccc|cc||cccc|cc}
\hline
& \multicolumn{6}{c||}{\textbf{SC-10}} & \multicolumn{6}{c}{\textbf{SC-30}} \\
\cline{2-13}
& \multicolumn{2}{c|}{\textbf{LSTM}} & \multicolumn{2}{c|}{\textbf{ResNet-18}} & \multicolumn{2}{c||}{\textbf{}} 
& \multicolumn{2}{c|}{\textbf{LSTM}} & \multicolumn{2}{c|}{\textbf{ResNet-18}} & \multicolumn{2}{c}{\textbf{}} \\
\cline{2-13}
& BA↑ & ASR↑ & BA↑ & ASR↑ & PESQ↑ & STOI↑  
& BA↑ & ASR↑ & BA↑ & ASR↑ & PESQ↑ & STOI↑ \\
\hline
PBSM       & \textbf{93.11} & 85.81 & 94.74 & 92.62 & 1.114 & 0.288 & 92.37 & 90.61 & 95.05 & 90.61 & 1.210 & 0.372 \\
JingleBack & 92.63 & 86.31 & 94.55 & 90.52 & 1.413 & 0.602 & 92.57 & 91.45 & 94.76 & 93.39 & 1.421 & 0.614 \\
Ultrasonic & 92.33 & 88.83 & 94.20 & \textbf{97.26} & 2.502 & 0.815 & \textbf{93.34} & 90.64 & 94.89 & \underline{96.44} & 2.892 & 0.845 \\
\hdashline
Bloodroot & \underline{92.75} & \textbf{95.83} & \textbf{95.01} & \underline{95.09} & \underline{3.002} & \underline{0.891} & \underline{93.18} & \textbf{96.82} & \underline{95.21} & \textbf{96.88} & \underline{3.031} & \underline{0.901} \\
Bloodroot-FT  & 92.44 & \underline{91.78} & \underline{94.82} & 93.85 & \textbf{3.315} & \textbf{0.915} & 92.48 & \underline{92.62} & \textbf{95.86} & 95.12 & \textbf{3.382} & \textbf{0.928} \\
\hline
\end{tabular}
\end{table*}

%%% Speaker Identification Task %%%%%%%%%%%%%%%%%%%%%%%%%%%%%%%%%%%%%%%%%%%%%%%%%%%%%%%
\begin{table*}[!htbp]
\centering
\caption{\textbf{Performance of the proposed Bloodroot-FT and baseline backdoor attacks on speaker identification (VoxCeleb-125 and VoxCeleb) at the 1\% poison rate}. 
BA: benign accuracy (\%); ASR: attack success rate (\%). Higher PESQ and STOI indicate better perceptual quality and intelligibility. The 
\textbf{Bold} style indicates the best performance, and the \underline{underlined} style indicates the second best.}
\label{tab:sid-results}
\renewcommand{\arraystretch}{1.1}
\setlength{\tabcolsep}{6.5pt}
\begin{tabular}{l|cccc|cc||cccc|cc}
\hline
& \multicolumn{6}{c||}{\textbf{VoxCeleb-125}} & \multicolumn{6}{c}{\textbf{VoxCeleb}} \\
\cline{2-13}
& \multicolumn{2}{c|}{\textbf{LSTM}} & \multicolumn{2}{c|}{\textbf{ResNet-18}} & \multicolumn{2}{c||}{\textbf{}} 
& \multicolumn{2}{c|}{\textbf{LSTM}} & \multicolumn{2}{c|}{\textbf{ResNet-18}} & \multicolumn{2}{c}{\textbf{}} \\
\cline{2-13}
& BA↑ & ASR↑ & BA↑ & ASR↑ & PESQ↑ & STOI↑  
& BA↑ & ASR↑ & BA↑ & ASR↑ & PESQ↑ & STOI↑ \\
\hline
PBSM       & 82.40 & 85.40 & \textbf{92.80} & 90.40 & 1.240 & 0.572 & \textbf{89.33} & 93.68 & \underline{91.37} & 95.28 & 1.245 & 0.573 \\
JingleBack & 81.60 & 65.60 & \underline{92.00} & 90.60 & 1.439 & 0.668 & 88.01 & 87.93 & \underline{91.37} & 94.12 & 1.426 & 0.659 \\
Ultrasonic & 80.80 & 86.41 & 91.20 & \underline{92.80} & 2.808 & 0.945 & \underline{88.73} & 94.88 & 91.13 & \underline{97.92} & 2.870 & 0.955 \\
\hdashline
Bloodroot & \underline{84.00} & \underline{89.60} & 91.20 & \textbf{97.60} & \underline{3.036} & \underline{0.975} & 88.57 & \textbf{98.24} & 91.29 & \underline{99.04} & \underline{3.079} & \underline{0.976} \\
Bloodroot-FT  & \textbf{85.60} & \textbf{89.60} & \underline{92.00} & \underline{96.00} & \textbf{3.327} & \textbf{0.977} & 88.37 & \underline{97.64} & \textbf{91.77} & \textbf{99.36} & \textbf{3.315} & \textbf{0.977} \\
\hline
\end{tabular}
\end{table*}

\vspace{-1em}
% \subsection{End-to-End Poison Refinement}
% \label{sec:method:finetune}
% We simulate the inference scenario with only 1\% poisoned data in the training process. The 1\% training data are relabeled to a targeted erroneous class while the remaining data are kept unchanged. 
% AudioSeal~\cite{audioseal} watermark entails a unique owner-to-identifier payload in the form of a 12-bit hashed message. We insert low-rank (LoRA)\cite{hu2022lora} adapters into the convolution layers of the decoder while other network layers are frozen.
\subsection{Watermark-Based Trigger Optimization}
\label{sec:method:optimization}
We consider a targeted threat where $1\%$ of training samples are poisoned via relabeling. To maximize attack effectiveness, we repurpose the AudioSeal \cite{audioseal} generator as a stealthy trigger source by inserting Low-Rank Adaptation (LoRA) \cite{hu2022lora} layers into its decoder blocks while freezing base parameters. This allows the pre-trained watermarker to adapt to the poisoning task with minimal overhead.

Formally, given clean audio $x \in \mathbb{R}^{B \times 1 \times T}$ ($B$ and $T$ as batch size and time), the poisonous trigger is $w_p = \alpha \cdot G(x)$. Here, $G(\cdot)$ is the generator and $\alpha$ is a scaling factor to balance attack potency against auditory transparency. We fine-tune $G$ via Eq. (5), which combines a supervised task loss to reinforce the trigger-label association with multiscale STFT and perceptual losses to ensure high acoustic fidelity.

\noindent\textbf{Supervised loss.} It encourages the generated watermark $\hat w = G(x)$ to match the targeted watermark $w_p$:
\begin{equation}
\mathcal{L}_{\mathrm{sup}}
= \left|\hat{w} - w_{p} \right|,
\end{equation}

\noindent\textbf{Multi-scale STFT loss.\cite{schwar2023multi}} It preserves the spectrogram similarity across multiple resolutions:
\begin{align}
\mathcal{L}_{\mathrm{stft}}
&= \frac{1}{N}\sum_{n=1}^{N}
\Big( | \hat{M}^{(n)}-M^{(n)} | + | \log_\varepsilon(\hat{M}^{(n)}) - \log_\varepsilon(M^{(n)}) | \Big)
\end{align}
where $\hat{M}^{(n)}$ and $M^{(n)}$ are predicted and original spectrograms for a given frequency resolution $n$, respectively, and $\log_\varepsilon(M)=\log(M+\varepsilon)$ uses a small offset $\varepsilon$ for stable log space calculation.

\noindent\textbf{Log-Mel perceptual loss.} It constrains the log-Mel deviations as follows, where $dB(M) = 10log_{10}M$:
\begin{equation}
\mathcal{L}_{\mathrm{mel}}
= \left| \mathrm{dB}\!\big(\hat{M}\big) - \mathrm{dB}\!\big(M\big) \right|.
\end{equation}

% \noindent\textbf{Multi-scale STFT loss.} It preserves the spectrogram similarity across multiple resolutions:
% \begin{align}
% \mathcal{L}_{\mathrm{stft}}
% &= \frac{1}{|\mathcal{N}|}\sum_{n\in\mathcal{N}}
% \Big( \| M_x^{(n)}-M_y^{(n)} \|_1 + \| \tilde M_x^{(n)}-\tilde M_y^{(n)} \|_1 \Big) \\
% M_x^{(n)}&=|\mathcal{S}_n(x)|\quad M_y^{(n)}=|\mathcal{S}_n(y)| \\
% \tilde M_x^{(n)}&=\log(M_x^{(n)}+\varepsilon)\quad \tilde M_y^{(n)}=\log(M_y^{(n)}+\varepsilon).
% \end{align}

% \noindent\textbf{Log-Mel perceptual loss.} It constrains the deviations in the Mel domain:
% \begin{equation}
% \mathcal{L}_{\mathrm{mel}}
% = \left\lVert \mathrm{dB}\!\big(\mathcal{M}(x)\big) - \mathrm{dB}\!\big(\mathcal{M}(y)\big) \right\rVert_{1},
% \end{equation}
% where $\mathcal{M}(\cdot)$ is the Mel-spectrogram transform.

% \noindent\textbf{Amplitude regularization.} It bounds the watermark energy to aid model convergence:
% \begin{equation}
% \mathcal{L}_{\mathrm{amp}}
% = \frac{1}{B\,T} \, \left\lVert \hat{w} \right\rVert_{2}^{2}.
% \end{equation}

\noindent\textbf{Amplitude regularization.} We introduce an amplitude penalty to prevent excessive perturbations that could degrade audio quality:
\begin{equation}
\mathcal{L}_{\mathrm{amp}} = \frac{1}{B\; T} \left\lVert \hat{w} \right\rVert_{2}^{2}.
\end{equation}
This constraint forces the generator to learn \textbf{structurally efficient} patterns rather than relying on brute-force energy increases. While $\alpha$ in Algorithm \ref{alg:poison} provides global scaling for potency, $\mathcal{L}_{\mathrm{amp}}$ ensures the fine-tuned triggers remain optimized for stealthiness within a bounded energy space.

\noindent\textbf{Total objective.}  
The overall loss function is defined as a weighted sum of all terms:  
\begin{equation}
\mathcal{L}
= \lambda_{\mathrm{sup}} \, \mathcal{L}_{\mathrm{sup}}
+ \lambda_{\mathrm{stft}} \, \mathcal{L}_{\mathrm{stft}}
+ \lambda_{\mathrm{mel}} \, \mathcal{L}_{\mathrm{mel}}
+ \lambda_{\mathrm{amp}} \, \mathcal{L}_{\mathrm{amp}}.
\end{equation}
After finetuning, we derive a poison generator $G_\alpha$ with a desired poison intensity specified by $\alpha$ and apply it to the downstream poisoning pipeline shown in Algorithm~\ref{alg:poison}.

% In our experiments, the weights are set as
% $\lambda_{\mathrm{sup}}=20000$,
% $\lambda_{\mathrm{stft}}=10$,
% $\lambda_{\mathrm{mel}}=10$,
% and $\lambda_{\mathrm{amp}}=0.1$. 
%We fine-tune only the LoRA parameters in the decoder while freezing the base generator. This design (i) focuses on the capacity of watermark synthesis, (ii) preserves the content modeling of the generator, and (iii) keeps the trainable footprint small. As a result, the refined model achieves stronger and more stable backdoor activation, with minimal PESQ / STOI degradation and reliable payload decoding for ownership attribution.

%The embedding strength $\alpha$ balances effectiveness and perceptual quality: A larger $\alpha$ improves learnability but lowers PESQ and short-time objective intelligibility (STOI), while a smaller $\alpha$ preserves quality but may reduce the ASR. In practice, $\alpha$ is set under PESQ / STOI thresholds to reflect perceptual budgets and maintain usability.

% \begin{figure*}[t]
%   \centering
%   % \includegraphics[width=0.8\textwidth, height=0.1\textheight]{fig/BA_ASR.png}
%   \includegraphics[width=1\textwidth]{fig/BA_ASR.png}
%   \caption{Impact of poisoning rate on SC-10. 
%   Shown are benign accuracy (BA) and attack success rate (ASR) under LSTM and ResNet-18 backbones. 
%   The figure illustrates the trade-off between attack effectiveness and model reliability across different poisoning levels.}
%   \label{fig:ba-asr}
% \end{figure*}

\begin{figure*}[t]
\centering

% ============ (a) BA vs Poisoning ============
\subfloat[BA vs.\ poisoning rate]{%
\begin{tikzpicture}
\begin{axis}[
  width=0.24\textwidth, height=0.18\textheight,
  xlabel={Poisoning rate (\%)}, ylabel={BA (\%)},
  xmin=0, xmax=2.0, ymin=92, ymax=96,
  xtick={0,0.5,1.0,1.5,2.0},
  grid=both, grid style={dashed,gray!40},
  tick style={semithick}, line width=0.9pt,
  legend to name=sharedlegend,
  legend style={draw=none,fill=none,legend columns=4, column sep=10pt}
]
% LSTM–AudioSeal
\addplot+[mark=diamond*, color=OIOrange, thick] coordinates
{ (0,93.68) (0.5,93.650) (1.0,92.754) (1.5,92.832) (2.0,92.871) };
\addlegendentry{LSTM–Bloodroot}
% LSTM–BloodRoot
\addplot+[mark=o, color=OIBlue, thick] coordinates
{ (0,93.68) (0.5,93.416) (1.0,92.44) (1.5,92.411) (2.0,92.053) };
\addlegendentry{LSTM–Bloodroot-FT}
% ResNet-18–AudioSeal
\addplot+[mark=diamond*, color=OIOrange!70!black, thick, dashed] coordinates
{ (0,95.11) (0.5,95.053) (1.0,95.014) (1.5,94.780) (2.0,94.235) };
\addlegendentry{ResNet-18–Bloodroot}
% ResNet-18–BloodRoot
\addplot+[mark=o, color=OIBlue!70!black, thick, dashed] coordinates
{ (0,95.11) (0.5,95.053) (1.0,94.820) (1.5,94.820) (2.0,94.820) };
\addlegendentry{ResNet-18–Bloodroot-FT}
\end{axis}
\end{tikzpicture}
\label{fig:ba_poison}}
\hfill
% ============ (b) ASR vs Poisoning ============
\subfloat[ASR vs.\ poisoning rate]{%
\begin{tikzpicture}
\begin{axis}[
  width=0.24\textwidth, height=0.18\textheight,
  xlabel={Poisoning rate (\%)}, ylabel={ASR (\%)},
  xmin=0, xmax=2.0, ymin=0, ymax=100,
  xtick={0,0.5,1.0,1.5,2.0},
  grid=both, grid style={dashed,gray!40},
  tick style={semithick}, line width=0.9pt,
  legend=false
]
% LSTM–AudioSeal
\addplot+[mark=diamond*, color=OIOrange, thick] coordinates
{ (0,0.0) (0.5,51.174) (1.0,95.826) (1.5,96.696) (2.0,96.870) };
% LSTM–BloodRoot
\addplot+[mark=o, color=OIBlue, thick] coordinates
{ (0,0.0) (0.5,59.000) (1.0,91.780) (1.5,92.783) (2.0,93.391) };
% ResNet-18–AudioSeal
\addplot+[mark=diamond*, color=OIOrange!70!black, thick, dashed] coordinates
{ (0,0.0) (0.5,77.826) (1.0,95.087) (1.5,95.565) (2.0,96.739) };
% ResNet-18–BloodRoot
\addplot+[mark=o, color=OIBlue!70!black, thick, dashed] coordinates
{ (0,0.0) (0.5,77.015) (1.0,93.850) (1.5,94.240) (2.0,95.391) };
\end{axis}
\end{tikzpicture}
\label{fig:asr_poison}}
\hfill
% ============ (c) LSTM PESQ–ASR ============
\subfloat[LSTM: PESQ vs.\ ASR]{%
\begin{tikzpicture}
\begin{axis}[
  width=0.24\textwidth, height=0.18\textheight,
  xlabel={PESQ}, ylabel={ASR (\%)},
  xmin=3.0, xmax=4.35, ymin=50, ymax=100,
  grid=both, grid style={dashed,gray!40},
  tick style={semithick}, line width=0.9pt,
  legend=false
]
% AudioSeal curve (5×→1×) + per-point labels
\addplot+[
  mark=*, color=OIOrange, thick,
  nodes near coords,
  point meta=explicit symbolic,
  every node near coord/.style={font=\scriptsize, anchor=west, xshift=3pt, yshift=3pt, text=OIOrange}
] coordinates {
  (3.031,96.82) [5×]
  (3.297,90.213) [4×]
  (3.561,83.215) [3×]
  (3.899,70.783) [2×]
  (4.281,53.412) [1×]
};
% BloodRoot point
\addplot+[only marks, mark=o, mark size=2.4pt, color=OIBlue, thick] coordinates { (3.315,92.62) };
\end{axis}
\end{tikzpicture}
\label{fig:lstm_pesq_asr}}
\hfill
% ============ (d) ResNet-18 PESQ–ASR ============
\subfloat[ResNet-18: PESQ vs.\ ASR]{%
\begin{tikzpicture}
\begin{axis}[
  width=0.24\textwidth, height=0.18\textheight,
  xlabel={PESQ}, ylabel={ASR (\%)},
  xmin=3.0, xmax=4.35, ymin=50, ymax=100,
  grid=both, grid style={dashed,gray!40},
  tick style={semithick}, line width=0.9pt,
  legend=false
]
% AudioSeal curve (5×→1×) + per-point labels
\addplot+[
  mark=*, color=OIOrange, thick,
  nodes near coords,
  point meta=explicit symbolic,
  every node near coord/.style={font=\scriptsize, anchor=west, xshift=3pt, yshift=3pt, text=OIOrange}
] coordinates {
  (3.031,96.879) [5×]
  (3.297,93.590) [4×]
  (3.561,88.018) [3×]
  (3.899,72.549) [2×]
  (4.281,54.705) [1×]
};
% BloodRoot point
\addplot+[only marks, mark=o, mark size=2.4pt, color=OIBlue, thick] coordinates { (3.315,95.12) };
\end{axis}
\end{tikzpicture}
\label{fig:res_pesq_asr}}
\\[2pt]
\pgfplotslegendfromname{sharedlegend}

\caption{Ablation study about the impact of the poisoning rate on SC-10. 
\textbf{(a)} Benign accuracy (BA) and \textbf{(b)} attack success rate (ASR). 
\textbf{(c)}–\textbf{(d)} PESQ–ASR trade-offs, illustrating how the poisoning rate affects both attack success and perceptual quality.}
\label{fig:four-panels}
\end{figure*}
\section{Experiments}
\label{sec:exp}
% \vspace{-2em}
\subsection{Setting}
\label{sec:exp:setting}

\noindent\textbf{Implementation details.}  
%\noindent\textbf{Training configuration.}  
The loss weights are $\lambda_{\mathrm{sup}}=20000$, 
$\lambda_{\mathrm{stft}}=10$, 
$\lambda_{\mathrm{mel}}=10$, 
and $\lambda_{\mathrm{amp}}=0.1$. 
All models are optimized using the Adam method with a learning rate of $1\times10^{-4}$ and a batch size of 32.
%\noindent\textbf{Bloodroot setting.}  
Bloodroot* applies Audioseal with $\alpha=5$ (no fine-tuning), while Bloodroot is the fine-tuned version using LoRA adapters.  
%\noindent\textbf{GPU resources.}  
Experiments are conducted using NVIDIA A16 for victim model training and A40 for Bloodroot fine-tuning.
%with LSTM and ResNet-18 backbones.

\noindent\textbf{Datasets.}  
For SR, we use SC-10 and SC-30, containing 10 and 30 keywords from the Speech Commands dataset~\cite{sc-dataset}, respectively.  
For SID, we use VoxCeleb-125 and the full VoxCeleb corpus~\cite{speakerid-voxceleb}, consisting of 125 speakers and the complete dataset.  
This setup enables examination of scalability across task types and class sizes.      

\noindent\textbf{Poisoning protocol.}  
We poison 1\% of training samples. 
In SR, the target is $y_t=\text{``left''}$; in SID, $y_t=\text{``id10020''}$.  
For label-flip poisoning, labels are reassigned to $y_t$. For clean-label poisoning, labels remain unchanged while the trigger is embedded. 
% Trigger strength is controlled to be comparable across methods.  

\noindent\textbf{Evaluation metrics.}  
We report \emph{benign accuracy (BA)} on clean inputs, \emph{attack success rate (ASR)} on triggered inputs, and perceptual quality via PESQ and short-time objective intelligibility (STOI)~\cite{STOI}.  

\noindent\textbf{Defenses.}  
We examine two defense mechanisms that are commonly employed during standard data and model processing workflows.  
\emph{Low-pass filtering} is a pre-processing defense using a 6th-order Butterworth filter with a cutoff frequency of $f_c=3800$ Hz, attenuating components above this threshold to remove high-frequency triggers (e.g., ultrasonic).  
\emph{Model pruning}\cite{anwar2017structured} is a post-training defense with \texttt{torch-pruning}, ranking convolutional channels by $L_2$ norm and pruning the least important ones along with dependent layers, yielding smaller yet consistent models.   

\begin{table}[t]
\caption{ASR before and after applying a spectral filtering defense. 
``Filter'' denotes a pre-processing filter applied before inference.}
\centering
\renewcommand{\arraystretch}{1.05}
\setlength{\tabcolsep}{6pt}
\begin{tabular}{lcc}
\hline
\textbf{Method} & \textbf{ASR (No Filter)} & \textbf{ASR (With Filter)} \\
\hline
PBSM              & 92.62\% & 9.52\% \\
JingleBack        & 90.52\% & 5.14\% \\
Ultrasonic        & 97.26\% & 1.28\% \\
Bloodroot       & 95.09\% & 44.58\% \\
Bloodroot-FT   & 93.85\% & \textbf{53.49\%} \\
\hline
\end{tabular}
\label{tab:asr-filter}
\end{table}

%先示意放上去
% \begin{figure}[t]
%   \centering
%   \includegraphics[width=0.95\columnwidth]{fig/pruning.png}
%   \caption{ASR performance under model pruning at different pruning rates. 
%   Timbre-based triggers show a sharp drop, while watermark-based triggers such as Bloodroot remain more stable.}
%   \label{fig:pruning}
% \end{figure}

% \subsection{Result}
% \label{sec:exp:result}

% --- 內文中插圖 ---
\begin{figure}[t]
\centering
\begin{tikzpicture}
\begin{axis}[
  width=0.95\columnwidth, height=0.18\textheight,
  xlabel={Pruning rate}, ylabel={ASR (\%)},
  xmin=0, xmax=1.0, ymin=0, ymax=100,         % <-- 壓縮 y 軸：20–100
  xtick={0,0.1,0.2,0.3,0.4,0.5,0.6,0.7,0.8,0.9,1.0},
  ytick={0,20,40,60,80,100},
  grid=both, grid style={dashed,gray!40},
  legend style={
    at={(0.5,1.02)}, anchor=south, legend columns=3,
    draw=none, fill=none, /tikz/every even column/.style={column sep=6pt}
  },
  line width=0.9pt, tick style={semithick},
]

% PBSM
\addplot+[mark=square*, color=OIGray, thick] coordinates {
(0.0,92.62) (0.1,0.304) (0.2,0.304) (0.3,0.304) (0.4,0.304)
(0.5,0.348) (0.6,0.348)
(0.7,0.261) (0.8,0.261) (0.9,0.261) (1.0,0.261)
};
\addlegendentry{PBSM}

% JingleBack
\addplot+[mark=triangle*, color=OIGreen, thick] coordinates {
(0.0,90.52) (0.1,1.696) (0.2,1.696) (0.3,1.739) (0.4,1.739)
(0.5,1.783) (0.6,1.783) (0.7,2.217) (0.8,2.217) (0.9,2.304) (1.0,2.304)
};
\addlegendentry{JingleBack}

% Ultrasonic
\addplot+[mark=pentagon*, color=OIPink, thick] coordinates {
(0.0,97.26) (0.1,0.261) (0.2,0.261) (0.3,0.261) (0.4,0.261)
(0.5,0.261) (0.6,0.261) (0.7,0.261) (0.8,0.261) (0.9,0.261) (1.0,0.261)
};
\addlegendentry{Ultrasonic}

% Bloodroot*
\addplot+[mark=diamond*, color=OIOrange, thick] coordinates {
(0.0,95.09) (0.1,28.870) (0.2,28.870) (0.3,29.304) (0.4,29.304)
(0.5,28.522) (0.6,28.522) (0.7,23.348) (0.8,23.348) (0.9,24.435) (1.0,24.435)
};
\addlegendentry{Bloodroot}

% BloodRoot
\addplot+[mark=o, color=OIBlue, thick] coordinates {
(0.0,93.85) (0.1,70.609) (0.2,70.609) (0.3,70.783) (0.4,70.783)
(0.5,70.522) (0.6,70.522) (0.7,64.565) (0.8,64.565) (0.9,64.391) (1.0,64.391)
};
\addlegendentry{Bloodroot-FT}

\end{axis}
\end{tikzpicture}
\vspace{-5pt}
\caption{ASR under a pruning defense\cite{anwar2017structured} across pruning rates.}
\label{fig:pruning}
\end{figure}

\vspace{-1em}

\subsection{Result}
\label{sec:exp:result}

From Tables~\ref{tab:sr-results} and~\ref{tab:sid-results}, both Bloodroot and Bloodroot-FT achieve higher BA and ASR on par compared to previous methods, but clearly surpass them in perceptual quality. Notably, Bloodroot improves PESQ by about 2 points and STOI by roughly 0.5 compared to other baselines. In contrast, methods such as PBSM and JingleBack often reduce perceptual quality or introduce audible artifacts, while the Ultrasonic trigger causes noticeable distortions.  

We further examine the effect of the poisoning rate on performance. As shown in Figs.~\ref{fig:ba_poison} and~\ref{fig:asr_poison}, BA remains largely stable as the poisoning rate increases, while ASR steadily improves. Even with as little as 0.5\% poisoned data, ASR already reaches about 50\%, demonstrating that the watermark trigger is effective across a wide range of poisoning levels. Beyond this, the PESQ--ASR trade-off curves (Figs.~\ref{fig:lstm_pesq_asr}, \ref{fig:res_pesq_asr}) show that Bloodroot consistently achieves higher ASR than Bloodroot at comparable PESQ values. This confirms that the LoRA fine-tuning step enhances both the effectiveness of the backdoor and its stealthiness, pushing the framework closer to the ideal balance of attack success and perceptual transparency.  

Since robustness against defenses is an essential evaluation criterion, we also test Bloodroot-FT under two widely used countermeasures: input filtering and model pruning (Table~\ref{tab:asr-filter}, Fig.~\ref{fig:pruning}). Results show that Bloodroot retains about 53\% ASR under low-pass filtering and around 70\% under pruning, indicating strong resilience. In contrast, non-watermark triggers have much worse performance, and their ASR values fall below 10\% and 5\%, respectively. The ultrasonic trigger, which depends on high-frequency perturbations, is almost entirely neutralized by filtering, and its ASR value is dropped to only 1.28\%. These results emphasize that watermark-based triggers not only sustain imperceptibility but also withstand common defense strategies more effectively than existing approaches.  

%Overall, Bloodroot matches prior work in terms of BA and ASR while offering significantly better perceptual quality and substantially stronger robustness to defenses. By combining effectiveness, imperceptibility, and resilience in a unified design, Bloodroot demonstrates that watermark-based triggers provide a practical and stealthy mechanism for backdoor attacks, and fine-tuning further extends their potential for real-world deployment.

%Against defenses, spectral filtering (Table~\ref{tab:asr-filter}) greatly lowers ASR for timbre-based triggers, but watermark-based ones stay robust with little loss. Under pruning (Fig.~\ref{fig:pruning}), performance is comparable to existing methods.

\vspace{-1em}
\vspace{-1em}
\section{Conclusion}
\vspace{-1em}
In this work, we present \textbf{Bloodroot}, the first \emph{watermark-as-trigger} framework for  \emph{audio data poisoning}. Bloodroot uses audio watermarking to build imperceptible and robust triggers. We further extend it to \textbf{Bloodroot-FT}, a fine-tuned version that improves both attack success and stealthiness. Experiments on SR and SID show that Bloodroot achieves higher perceptual quality (PESQ/STOI) while maintaining competitive ASR and BA. It remains robust under filtering and pruning. Together, Bloodroot and Bloodroot-FT demonstrate that watermark-based triggers can serve as a practical and stealthy backdoor mechanism. They achieve the goals of effectiveness, imperceptibility, and accountability. Looking forward, the inherent attribution property of watermarks may offer further extensions for backdoor design and analysis, potentially expanding to other downstream tasks such as speech emotion recognition \cite{SER-1, SER-2}.

\bibliographystyle{IEEEbib}
\bibliography{mybib}

@article{liu2025xattnmark,
  title={XAttnMark: Learning robust audio watermarking with cross-attention},
  author={Liu, Yixin and Lu, Lie and Jin, Jihui and Sun, Lichao and Fanelli, Andrea},
  journal={arXiv preprint arXiv:2502.04230},
  year={2025}
}

@inproceedings{audioseal,
  TITLE = {{Proactive detection of voice cloning with localized watermarking}},
  AUTHOR = {Roman, Robin San and Fernandez, Pierre and Elsahar, Hady and D{\'e}fossez, Alexandre and Furon, Teddy and Tran, Tuan},
  URL = {https://hal.science/hal-04610152},
  BOOKTITLE = {{Proceedings of the 41st International Conference on Machine Learning}},
  ADDRESS = {Vienna, Austria},
  ORGANIZATION = {{PMLR}},
  VOLUME = {235},
  PAGES = {1-17},
  YEAR = {2024},
  MONTH = Jul,
  HAL_ID = {hal-04610152},
  HAL_VERSION = {v1},
}

@article{timbreMark,
  title={Detecting voice cloning attacks via timbre watermarking},
  author={Liu, Chang and Zhang, Jie and Zhang, Tianwei and Yang, Xi and Zhang, Weiming and Yu, Nenghai},
  journal={arXiv preprint arXiv:2312.03410},
  year={2023}
}

@article{wavmark,
  title={Wavmark: Watermarking for audio generation},
  author={Chen, Guangyu and Wu, Yu and Liu, Shujie and Liu, Tao and Du, Xiaoyong and Wei, Furu},
  journal={arXiv preprint arXiv:2308.12770},
  year={2023}
}

@INPROCEEDINGS{Dataownership,
  author={Bouaziz, Wassim Wes and El-Mhamdi, El-Mahdi and Usunier, Nicolas},
  booktitle={ICASSP 2025 - 2025 IEEE International Conference on Acoustics, Speech and Signal Processing (ICASSP)}, 
  title={Targeted Data Poisoning for Black-Box Audio Datasets Ownership Verification}, 
  year={2025},
  volume={},
  number={},
  pages={1-5},
  doi={10.1109/ICASSP49660.2025.10888515}}

@INPROCEEDINGS{Watermark-poison,
  author={Tang, Yi},
  booktitle={ICASSP 2025 - 2025 IEEE International Conference on Acoustics, Speech and Signal Processing (ICASSP)}, 
  title={Poisoning The Diffusion: A Simple and Robust Watermarking Method for Audio Generation}, 
  year={2025},
  volume={},
  number={},
  pages={1-5},
  doi={10.1109/ICASSP49660.2025.10889187}}

@inproceedings{TraceableSpeech,
  title     = {TraceableSpeech: Towards Proactively Traceable Text-to-Speech with Watermarking},
  author    = {Junzuo Zhou and Jiangyan Yi and Tao Wang and Jianhua Tao and Ye Bai and Chu Yuan Zhang and Yong Ren and Zhengqi Wen},
  year      = {2024},
  booktitle = {Interspeech 2024},
  pages     = {2250--2254},
  doi       = {10.21437/Interspeech.2024-534},
  issn      = {2958-1796},
}

@article{Encodec,
  title={High Fidelity Neural Audio Compression},
  author={Défossez, Alexandre and Copet, Jade and Synnaeve, Gabriel and Adi, Yossi},
  journal={arXiv preprint arXiv:2210.13438},
  year={2022}
}

@inproceedings{
dynamic-superb,
title={Dynamic-{SUPERB} Phase-2: A Collaboratively Expanding Benchmark for Measuring the Capabilities of Spoken Language Models with 180 Tasks},
author={Chien-yu Huang and others},
booktitle={The Thirteenth International Conference on Learning Representations},
year={2025},
url={https://openreview.net/forum?id=s7lzZpAW7T}
}

@article{backdoor-survey,
  title={Backdoor learning: A survey},
  author={Li, Yiming and Jiang, Yong and Li, Zhifeng and Xia, Shu-Tao},
  journal={IEEE transactions on neural networks and learning systems},
  volume={35},
  number={1},
  pages={5--22},
  year={2022},
  publisher={IEEE}
}

@article{Backdoor-stealthy,
  title={Toward stealthy backdoor attacks against speech recognition via elements of sound},
  author={Cai, Hanbo and Zhang, Pengcheng and Dong, Hai and Xiao, Yan and Koffas, Stefanos and Li, Yiming},
  journal={IEEE Transactions on Information Forensics and Security},
  volume={19},
  pages={5852--5866},
  year={2024},
  publisher={IEEE}
}

@inproceedings{Backdoor-ACM,
  title={Opportunistic backdoor attacks: Exploring human-imperceptible vulnerabilities on speech recognition systems},
  author={Liu, Qiang and Zhou, Tongqing and Cai, Zhiping and Tang, Yonghao},
  booktitle={Proceedings of the 30th ACM International Conference on Multimedia},
  pages={2390--2398},
  year={2022}
}

@inproceedings{Ultrasonic,
  author = {Koffas, Stefanos and Xu, Jing and Conti, Mauro and Picek, Stjepan},
  title = {Can You Hear It? Backdoor Attacks via Ultrasonic Triggers},
  year = {2022},
  isbn = {9781450392778},
  publisher = {Association for Computing Machinery},
  address = {New York, NY, USA},
  url = {https://doi.org/10.1145/3522783.3529523},
  doi = {10.1145/3522783.3529523},
  booktitle = {Proceedings of the 2022 ACM Workshop on Wireless Security and
  Machine Learning},
  pages = {57–62},
  numpages = {6},
  location = {San Antonio, TX, USA},
  series = {WiseML '22}
}

@inproceedings{JingleBack,
  title={Going In Style: Audio Backdoors Through Stylistic Transformations},
  author={Koffas, Stefanos and Pajola, Luca and Picek, Stjepan and Conti, Mauro},
  booktitle={ICASSP 2023-2023 IEEE International Conference on Acoustics, Speech and Signal Processing (ICASSP)},
  pages={1--5},
  year={2023},
  organization={IEEE}
}

@inproceedings{SER-1,
  title     = {{CLEP-DG: Contrastive Learning for Speech Emotion Domain Generalization via Soft Prompt Tuning}},
  author    = {Jiacheng Shi and others},
  year      = {2025},
  booktitle = {{Interspeech 2025}},
}

@INPROCEEDINGS{SER-2,
  author={Lin, Hsi-Che and Lin, Yi-Cheng and Chou, Huang-Cheng and Lee, Hung-yi},
  booktitle={ICASSP 2025 - 2025 IEEE International Conference on Acoustics, Speech and Signal Processing (ICASSP)}, 
  title={Improving Speech Emotion Recognition in Under-Resourced Languages via Speech-to-Speech Translation with Bootstrapping Data Selection}, 
  year={2025},
}

@inproceedings{CBA,
  title     = {{CBA: Backdoor attack on deep speech classification via audio compression}},
  author    = {Yuheng Huang and Ying Ren and Wenjie Zhang and Diqun Yan},
  year      = {2025},
  booktitle = {{Interspeech 2025}},
  pages     = {5648--5652},
  doi       = {10.21437/Interspeech.2025-372},
  issn      = {2958-1796},
}

@inproceedings{voxpopuli,
    title = "{V}ox{P}opuli: A Large-Scale Multilingual Speech Corpus for Representation Learning, Semi-Supervised Learning and Interpretation",
    author = "Wang, Changhan  and
      Riviere, Morgane  and
      Lee, Ann  and
      Wu, Anne  and
      Talnikar, Chaitanya  and
      Haziza, Daniel  and
      Williamson, Mary  and
      Pino, Juan  and
      Dupoux, Emmanuel",
    booktitle = "Proc. 59th Annual Meeting of the Association for Computational Linguistics and the 11th Int. Joint Conf. Natural Language Processing (Volume 1: Long Papers)",
    month = aug,
    year = "2021",
    publisher = "Association for Computational Linguistics",
    url = "https://aclanthology.org/2021.acl-long.80",
    pages = "993--1003",
}

@article{sc-dataset,
  title={Speech commands: A dataset for limited-vocabulary speech recognition},
  author={Warden, Pete},
  journal={arXiv preprint arXiv:1804.03209},
  year={2018}
}

@INPROCEEDINGS{clean-label,
  author={Xinyuan, Henry Li and Joshi, Sonal and Thebaud, Thomas and Villalba, Jesus and Dehak, Najim and Khudanpur, Sanjeev},
  booktitle={2024 IEEE Spoken Language Technology Workshop (SLT)}, 
  title={Clean Label Attacks Against SLU Systems}, 
  year={2024},
  volume={},
  number={},
  pages={1107-1114},
  doi={10.1109/SLT61566.2024.10832144}}

@article{poison-sr,
  title={Query-efficient adversarial attack with low perturbation against end-to-end speech recognition systems},
  author={Wang, Shen and Zhang, Zhaoyang and Zhu, Guopu and Zhang, Xinpeng and Zhou, Yicong and Huang, Jiwu},
  journal={IEEE Transactions on Information Forensics and Security},
  volume={18},
  pages={351--364},
  year={2022},
  publisher={IEEE}
}

@article{speakerid-voxceleb,
  title={Voxceleb: A large-scale speaker identification dataset},
  author={Nagrani, Arsha and Chung, Joon Son and Zisserman, Andrew},
  journal={arXiv preprint arXiv:1706.08612},
  year={2017}
}

@article{poison-speakerid,
  title={Dictionary attacks on speaker verification},
  author={Marras, Mirko and Korus, Pawe{\l} and Jain, Anubhav and Memon, Nasir},
  journal={IEEE Transactions on Information Forensics and Security},
  volume={18},
  pages={773--788},
  year={2022},
  publisher={IEEE}
}

@ARTICLE{STOI,
  author={Taal, Cees H. and Hendriks, Richard C. and Heusdens, Richard and Jensen, Jesper},
  journal={IEEE Transactions on Audio, Speech, and Language Processing}, 
  title={An Algorithm for Intelligibility Prediction of Time–Frequency Weighted Noisy Speech}, 
  year={2011},
  volume={19},
  number={7},
  pages={2125-2136},
  doi={10.1109/TASL.2011.2114881}}

@article{anwar2017structured,
  title={Structured pruning of deep convolutional neural networks},
  author={Anwar, Sajid and Hwang, Kyuyeon and Sung, Wonyong},
  journal={ACM Journal on Emerging Technologies in Computing Systems (JETC)},
  volume={13},
  number={3},
  pages={1--18},
  year={2017},
  publisher={ACM New York, NY, USA}
}

@article{schwar2023multi,
  title={Multi-scale spectral loss revisited},
  author={Schw{\"a}r, Simon and M{\"u}ller, Meinard},
  journal={IEEE Signal Processing Letters},
  volume={30},
  pages={1712--1716},
  year={2023},
  publisher={IEEE}
}

@article{hu2022lora,
  title={Lora: Low-rank adaptation of large language models.},
  author={Hu, Edward J and Shen, Yelong and Wallis, Phillip and Allen-Zhu, Zeyuan and Li, Yuanzhi and Wang, Shean and Wang, Lu and Chen, Weizhu and others},
  journal={ICLR},
  volume={1},
  number={2},
  pages={3},
  year={2022}
}

\end{document}